\documentclass[12pt]{article}
\usepackage{bm}
\usepackage{float}
\usepackage{graphicx, mathrsfs,amssymb,amsmath,amsfonts,amsbsy,amsthm,eufrak,latexsym}
\begin{document}

\newcommand{\Dh}{\hbox{\bf D}}
\newcommand{\Xh}{\hbox{\bf X}}
\newcommand{\Fh}{\hbox{\bf F}}
\newcommand{\nablaslh}{\nabla\raise2pt\hbox{\kern-9pt\slash\kern3pt}}

\title{\vskip.5cm
Lattice calculations on the spectrum of Dirac and Dirac-K\"ahler operators
\vskip.5cm}
\author{R.G. Campos, J.L. L\'opez-L\'opez, R. Vera\\
Facultad de Ciencias F\'{\i}sico-Matem\'aticas, \\
Universidad Michoacana, \\
58060, Morelia, Mich., M\'exico.\\
\hbox{\small rcampos@umich.mx, jorge@fismat.umich.mx, rvera@umich.mx}\\
}
\date{}
\maketitle
{\vskip.3cm
\noindent MSC: 58J05, 65D25, 65T40, 81Q05
}\\
\vspace*{1.5truecm}
\begin{center} Abstract \end{center}
We present a matrix technique to obtain the spectrum and the analytical index of some elliptic operators defined on compact Riemannian manifolds. The method uses matrix representations of the derivative which yield exact values for the derivative of a trigonometric polynomial. These matrices can be used to find the exact spectrum of an elliptic operator in particular cases and in general, to give insight into the properties of the solution of the spectral problem. As examples, the analytical index and the eigenvalues of the Dirac operator on the torus and on the sphere are obtained and as an application of this technique, the spectrum of the Dirac-K\"ahler operator on the sphere is explored. 
\vskip1.5cm
\section{Introduction}\label{intro}
The problem of finding the spectrum and eigenstates of elliptic operators on compact Riemannian manifolds is of considerable interest in physical theories, from quantum field theory to gravitation and cosmology. The spectral problem of the Dirac operator on the sphere, for instance, is related to confinement of quarks; it has been studied since many years ago and has been solved on the $n$-dimensional sphere \cite{Tra95, Por95, Abr02, Cba96}. The spectral problem of the so-called Dirac-K\"ahler operator acting on differential forms, postulated  by Ivanenko and Landau in 1928 as an operator to describe properties of fermionic fields and connected with the Dirac equation by K\"ahler in 1962,\footnote{In fact, the Dirac-K\"ahler equation is equivalent to several independent simultaneous Dirac equations in a flat manifold \cite{Obu94}.} is also physically relevant since it has been used to describe differential forms as spinors at the time that it exhibits bosonic-like solutions \cite{Obu94, Kru02, Ben83, Gra78}, giving an alternative framework for lattice theory \cite{Bec82, Mon94, Kan04, Bea03}.\\
In a sequel of papers (\cite{Cam02}-\cite{Cam00b}, and references therein), a nonlocal formulation of lattice theory has been introduced. In some simple cases this technique yields exact results, i.e., the numerical output can be interpolated to obtain the functions that solve exactly the problem. In this method, the discretization of a differential operator is obtained by the substitution of the derivatives appearing in the operator by matrix representations of the derivative which arise naturally in the context of interpolation of functions.  Once the finite-dimensional operator is constructed, it can be solved by standard techniques.\\
The aim of this paper is to show the potential use of this lattice formulation by obtaining the spectrum and analytical index of the Dirac equation on the torus and on the sphere and exploring the eigenvalue problem of the Dirac-K\"ahler operator on these surfaces. \\
We solve the Dirac operator on the sphere and find that besides the known fermionic spectral solution with eigenvalues $\pm l$, $l=1,2,\ldots$, there are present a bosonic-like solution of the spectral problem with eigenvalues $0, \pm 3/2, \pm 5/2, \ldots$ \\
Concerning the Dirac-K\"ahler spectral problem, we find that the spectrum splits into two sets: one corresponding to the bosonic-like eigenvalues $\pm\sqrt{l(l+1)}$, $l=0,1,\ldots$, and the other to a fermionic-like counterpart consisting of numbers approaching the integer eigenvalues of the Dirac operator on the sphere with double multiplicity, confirming that in general, the Dirac-K\"ahler operator cannot be decomposed in several (two in this case) independent Dirac equations on a curved manifold.\\
It is worth to notice that the use of this technique allows to find analytically the Dirac spectrum on a multidimensional torus.
\section{Differentiation matrices} \label{secdifmat}
In this section we present the main results on the discretization method based on the use of differentiation matrices. Proofs and further applications can be found in  \cite{Cam04, Cam00b}.\\
Let us consider $N=2n+1$ arbitrary points $-\pi<x_1<x_2<\cdots<x_{2n+1}\le\pi$ and the $N\times N$ matrix defined by
\begin{equation}\label{difmattrigpol}
D=T\tilde{D}T^{-1},
\end{equation}
where $\tilde{D}$ and $T$ are the matrices whose entries are given by
\begin{equation}\label{ddij}
\tilde{D}_{ij}=\begin{cases}\displaystyle\mathop{{\sum}}\limits_{l\ne i}^N
{1\over2}\cot{{(x_i-x_l)}\over 2},&{i=j},\cr\noalign{\vskip .5truecm}
\displaystyle {1\over2}\csc{{(x_i-x_j)}\over 2}, &{i\not=j},\cr\end{cases}
\hskip 1.7truecm
\displaystyle T_{ij}=t'(x_i)\delta_{ij},\qquad i,j=1,\ldots N,
\end{equation}
where 
\[
t'(x_i)={d\over{dx}}\Big[\prod_{l=1}^N\sin{{x-x_l}\over 2}\Big]_{x_i}.
\]
If $f(x)$ is a trigonometric polynomial of degree at most $n$ and 
$f^{(k)}$ denote the vector whose elements are the derivatives $f^{(k)}(x_j)$, $j=1,\ldots,N$, then
\begin{equation}\label{fdk}
f^{(k)}=D^k f, \quad k=0,1,\ldots
\end{equation}
If an even number of nodes, say $N=2n$, is selected instead of an odd number, this equation still holds but for antiperiodic trigonometric polynomials, i.e., functions of the form $f(x)=\exp(ix/2)g(x)$ where $g(x)$ is a trigonometric polynomial of degree at most $n-1$.\\
If the periodic function $f(x)\in L_2(-\pi,\pi)$ is not a polynomial, a residual vector must be added to the right-hand side of (\ref{fdk}). The norm of this vector approaches zero as the number of nodes is increased since $f(x)$ can be expanded in a Fourier series.\\
The differentiation matrix $D$ takes the simple form 
\begin{equation}\label{djk}
D_{jk}=\begin{cases} 0,&{j=k},\cr\noalign{\vskip .5truecm}
\displaystyle {(-1)^{j+k}\over{2\sin{\pi\over N}(j-k) }}, &{j\not=k},\cr\end{cases}
\end{equation}
when the nodes are chosen as the evenly spaced points 
\begin{equation}
x_j={\pi\over N} (2j-N-1), \quad j=1,2,\ldots,N.
\label{tj}
\end{equation}
The generalization to the case of $q$ variables $x^1,x^2,\ldots,x^q$ is straightforward. The space of functions where the differentiation matrices become the partial derivatives is the tensor product of the subspaces of trigonometric polynomials of degree at most $n_k$ in $x^k$. Let $N_k$ be the number of nodes in the direction $k$. Then, the total number of nodes is $N=\prod_{k=1}^q N_k$ and they are indexed  by
\begin{equation}\label{ordenlatpun}
r=j_1+(j_2-1)N_1+(j_3-1)N_1N_2+\cdots+(j_q-1)\prod_{k=1}^{q-1}N_k,
\end{equation}
where $j_k=1,2,\ldots,N_k$, $k=1,2,\ldots, q$.
The $N\times N$ differentiation matrix representing 
$\partial^k/\partial x^k$, is now
\begin{equation}\label{eqDbfk}
{\bf D}_k=1_{N_q}\otimes\cdots\otimes 1_{N_{k+1}}\otimes D_k\otimes 1_{N_{k-1}}\otimes\cdots\otimes 1_{N_1},
\end{equation}
where $k=1,2,\ldots, q$ and it should be multiplied by the vector {\bf f} whose entries are
\[
f_r=f(x^1_{j_1},x^2_{j_2},\cdots,x^q_{j_q}), 
\]
in order to obtain $\partial f(x^1_{j_1},x^2_{j_2},\cdots,x^q_{j_q})/\partial  x^k$.\\
The differentiation matrices (\ref{eqDbfk}) can be  diagonalized simultaneously by the unitary and 
symmetric matrix 
\begin{equation} \label{seis}
{\mathbf F}=F_q\otimes\cdots\otimes F_1.
\end{equation}
Each $N_k\times N_k$ matrix $F_k$ is a discrete Fourier transform with the property
\[
(F_k^\dagger D_k F_k)_{jm}=id^{(k)}_m\delta_{jm}.
\]
The components of $F_k$ are given by
\begin{equation}\label{eqFtrigpol}
(F_k)_{jm}=\frac{1}{N_k}\displaystyle{e^{-2\pi i jd^{(k)}_m/N_k}},
\end{equation}
and the eigenvalues of $D_k$ are 
\begin{equation}\label{eqsetvaloi}
d^{(k)}_m=-(N_k+1)/2+m,
\end{equation}
where $j,m=1,\ldots, N_k$.
\section{Lattice Dirac operators}\label{secDirops}
In this section, we show how the eigenvalues, the eigenspinors and the analytical index of the free Dirac operator on the flat torus and on the sphere can be obtained from a discretized version of the Dirac operator constructed with differentiation matrices. 
\subsection{The flat torus}\label{subsecDirtor}
A two-dimensional flat torus is a Riemannian quotient manifold $T_\Lambda=\mathbb R^2/\Lambda$, where $\Lambda$ is a discrete group generated by the translations on the directions $\lambda_1=(1,0)$, $\lambda_2=(a,b)$, where $b>0$.  There are four spin structures on $T_\Lambda$, each of them is characterized by $(\mu_1, \mu_2)$, where $\mu_{1,2}=0,1$ and has associated a spin bundle whose sections are spinor fields. A spinor field is a function $f: \mathbb R^2\to \mathbb C^2$ which is periodic or antiperiodic in the $\lambda_k$-direction if $\mu_k$ is 0 or 1 respectively.\\ 
In order to use differentiation matrices it is required a $2\pi$-periodicity in the orthogonal coordinates $x$ and $y$ instead of the natural $\Lambda$-periodicity of the spinor fields. Since the linear function $\psi(x,y)=(x+ay,by)/2\pi$ in $\mathbb R^2$ yields a diffeomorphism between the torus $\mathbb R^2/(2\pi \mathbb Z)^2$ and $T_\Lambda$, we can use this change of variable to write the Dirac operator on $T_\Lambda$ in terms of the coordinates $x$ and $y$. Thus, the Dirac operator takes the form 
\begin{equation}\label{Diroptor}
\begin{pmatrix} 0 & D^-\\  D^+ & 0\end{pmatrix},
\end{equation}
where 
\begin{equation}\label{Diroptorpm}
2\pi D^\pm=(i\pm\dfrac{a}{b}) \dfrac{\partial}{\partial x}\mp\dfrac{1}{b}\dfrac{\partial}{\partial y}.
\end{equation}
In general, the spectrum of the Dirac operator depends on the choice of the spin structure. For $T_\Lambda$ the spectrum of $D$ is already known \cite{Fri84}. If the spin structure of $T_\Lambda$ is given by $(\mu_1,\mu_2)$ the eigenvalues of (\ref{Diroptor}) have the form
\begin{equation}\label{eigdiroptor}
\pm \left\Vert (m+ \frac{\mu_1}{2})\lambda^*_1+(n+ \frac{\mu_2}{2}) \lambda^*_2 \right\Vert
\end{equation}
where $\lambda^*_1=(1,-a/b)$, $\lambda^*_2=(0,1/b)$ and $\Vert\cdot\Vert$ stands for the euclidean norm in $\mathbb R^2$. Each pair of integers $m$ and $n$ 
yields one eigenvalue of multiplicity $K$, where $K$ is the number of possible combinations of $m$ and $n$ giving the same eigenvalue.\\
The use of differentiation matrices yields a discrete problem with the same result as we show in the following. Since the eigenspinors are $2\pi$-periodic functions in the $k$-direction if $\mu_k=0$ and $2\pi$-antiperiodic functions if $\mu_k=1$, we can use the differentiation matrix (\ref{djk}) to yield the discretized form of the Dirac operator (\ref{Diroptor})  
\begin{equation}\label{diropress}
{\mathbf D}=\begin{pmatrix} 0 & {\mathbf D}^-\\  {\mathbf D}^+ & 0\end{pmatrix},
\end{equation}
where 
\begin{equation}\label{diroptordis}
{\mathbf D}^\pm=(i\pm\dfrac{a}{b}) \mathbf D_ 1\mp\dfrac{1}{b} \mathbf D_ 2,
\end{equation}
${\mathbf D}_k$ is given by (\ref{eqDbfk}). The spin structure will determine whether the number $N_k$ of nodes is odd or even.\\
The eigenvalue problem of the discretized Dirac operator ${\mathbf D}$ can be solved by noting that
\begin{equation}\label{detdiropr}
\det({\mathbf D}-\lambda {\mathbf 1})=\det\left(\lambda^2{\mathbf 1}-(1+\dfrac{a^2}{b^2}){\mathbf D}_1^2-\dfrac{1}{b^2}{\mathbf D}_2^2+\dfrac{2a}{b^2}{\mathbf D}_1{\mathbf D}_2\right),
\end{equation}
where we have used the formula of Schur and the fact that ${\mathbf D}_k$ commutes with ${\mathbf D}_j$. This matrix can be diagonalized by the 
matrix $\mathbf F$, where ${\mathbf F}$ is the two-dimensional discrete Fourier transform given by (\ref{seis}). Taking into account the structure of (\ref{detdiropr}), we get that the eigenvalues of the discretized Dirac operator are of the form
\begin{equation}\label{eigopdirmet}
\pm \sqrt{(1+\dfrac{a^2}{b^2})(d^{(1)})^2+\dfrac{1}{b^2}(d^{(2)})^2-\dfrac{2a}{b^2}d^{(1)}d^{(2)}}=\pm \Vert (i+\dfrac{a}{b})d^{(1)}- \dfrac{1}{b}d^{(2)}\Vert_2
\end{equation}
where $d^{(k)}$ is one of the eigenvalues of $D_k$ given by (\ref{eqsetvaloi}) and $\Vert\cdot\Vert_2$ stands for the usual hermitian norm in $\mathbb C^2$. If the spin structure is not twisted (the trivial structure), the eigenspinors are $2\pi$-periodic trigonometric polynomials and $N_k$ must be an odd number. Therefore, 
$d^{(k)}$ runs over ${\mathbb Z}$ when $N_k\to\infty$ for $k=1,\ldots,n$ [cf. (\ref{eqsetvaloi})]. If the spin structure is twisted in the $k$-direction, the corresponding component of the eigenspinors is $2\pi$-antiperiodic, $N_k$ must be now an even number and $d^{(k)}$ has the form $l+1/2$, where $l$ is an integer. Thus, in the asymptotic limit $N_k\to\infty$,  (\ref{eigopdirmet}) coincides with (\ref{eigdiroptor}). Note the relation between the spin structure $(\mu_1,\mu_2)$ and the parity of the set $(N_1,N_2)$. The zero eigenvalue appears only in the trivial structure.\\
The analytical index $n_D$ of the Dirac operator is defined by
\[
n_D=\dim\ker D^+-\dim\ker D^-.
\] 
Therefore, the  finite-dimensional version of $n_D$ is the difference
\begin{equation}\label{discindx}
n^*_D=\dim\ker {\mathbf D}^+-\dim\ker {\mathbf D}^-, 
\end{equation}
where ${\mathbf D}^\pm$ are the block matrices (\ref{diroptordis}). Since the lattice Dirac operator is invertible, as the continuum operator, the index $n^*_D$ is zero. However, we can compute explicitly $n^*_D$ by counting the null eigenvalues of ${\mathbf D}^\pm$. Again, these matrices can be diagonalized by $\mathbf F$. In this way we find that the eigenvalues of ${\mathbf D}^+$ have the form
\begin{equation}\label{eigdiropup}
(i+\dfrac{a}{b})\,d^{(1)}- \dfrac{1}{b}\,d^{(2)}
\end{equation}
and those of ${\mathbf D}^-$ have the form
\begin{equation}\label{eigdiropdown}
(i-\dfrac{a}{b})\,d^{(1)}+ \dfrac{1}{b}\,d^{(2)}.
\end{equation}
Therefore, both ${\mathbf D}^+$ and ${\mathbf D}^-$ have null eigenvalues only in the case of the trivial structure, because only in that case, $d^{(1,2)}$ can be zero. From (\ref{eigdiropup}) and (\ref{eigdiropdown}), we can see that the number of null eigenvalues of ${\mathbf D}^+$ is the same number of null eigenvalues of ${\mathbf D}^-$. Therefore $n^*_D=n_D$, i.e., the analytical index of the discrete free Dirac operator on $T_\Lambda$ is zero and this only occurs for the trivial spin structure.\\
The above procedure can be straightforwardly generalized to higher-dimensional tori. 
\subsection{The Dirac operator on the sphere}
This is another example of application of differentiation matrices yielding exact results. 
Let us consider the standard parametrization $\phi: (0,\pi)\times(0,2\pi)\to{\mathbb R}^3$ of the unitary sphere $S^2$ given by
\[
\phi(\theta,\varphi)=(\sin\theta\cos\varphi, \sin\theta\sin\varphi, \cos\theta).
\]
The hermitian Dirac operator takes the form
\begin{equation}\label{diropshp}
D=\begin{pmatrix} 0&&i \dfrac{\partial}{\partial\theta}+\dfrac{i}{2}\cot\theta+\csc\theta \dfrac{\partial}{\partial\varphi}\\
i \dfrac{\partial}{\partial\theta}+\dfrac{i}{2}\cot\theta-\csc\theta \dfrac{\partial}{\partial\varphi}&&0\end{pmatrix}.
\end{equation}
This problem has been solved in higher dimensions \cite{Tra95, Por95, Cba96}. In this case the eigenvalues are $\pm 1,\pm 2,\ldots,\pm n\ldots$ with multiplicity $2n$. The eigenspinors are given in terms of products of $2\pi$-antiperiodic trigonometric polynomials in $\theta$ and $\varphi$. To derive a discrete Dirac operator we have to take into account that $0<\theta<\pi$ and therefore, the differentiation matrix $D_\theta$ can not be constructed according to (\ref{djk}). However, we can use the general form (\ref{difmattrigpol}) instead of  (\ref{djk}) and choose $N_\theta$ different points in $(0,\pi)$, or even better, take the unique solution 
$0<\theta_1<\theta_2<\cdots<\theta_{N_\theta}<\pi$ of
\begin{equation}\label{eqnodt}
\sum_{k\ne j}^{N_\theta}\cot{{\theta_j-\theta_k}\over 2}=-\frac{1}{2}\cot\theta_j
\end{equation}
as nodes \cite{Cam00b}. This selection yields a hermitian discrete Dirac operator ${\bf D}$ of the form (\ref{diropress}) where now ${\mathbf D}^\pm$ are given by
\begin{equation}\label{dpmsph}
{\mathbf D}^\pm=i1_\varphi\otimes D_\theta\pm D_\varphi\otimes\csc\Theta ,
\end{equation}
where $\Theta$ is a diagonal matrix whose nonzero elements are the numbers $\theta_k$. The matrix corresponding to the term $(\cot\theta)/2$ of (\ref{diropshp}) does not appear here because of (\ref{eqnodt}) and therefore, $D_\theta$ has null diagonal elements and it is a differentiation matrix for the operator
\[
\frac{1}{\sqrt{\sin\theta}}\dfrac{d}{d\theta}\sqrt{\sin\theta}
\]
acting on trigonometric polynomials. The differentiation matrix $D_\varphi$ may be constructed according to (\ref{djk}) since there is no restriction on $\varphi$. Note that the number of nodes to be used in the construction of the differentiation matrices must be even integers in both directions. The eigenproblem ${\bf D}{\bm\psi}_n=\lambda_n{\bm\psi}_n$ can be solved by the use of standard routines to
produce the right results: if $\min(N_\theta, N_\varphi)$ is denoted by $N$, the exact eigenvalues 
\[  
\pm 1,\pm 2,\cdots, \pm k,\cdots,\pm (N-2)/2  
\]
appear among the eigenvalues of ${\bf D}$ with the right multiplicity $2k$. In the limit $N\to\infty$, the complete solution is recovered. The eigenvectors can be interpolated to give the correct eigenspinors but it is necessary to take into account that this is a degenerate eigenvalue problem. For example, let us take the eigenvalue $\lambda_1=1$. The corresponding eigenspinors
\[
\psi_1(\theta,\varphi)=\begin{pmatrix}e^{i\varphi/2} \cos\theta/2\\-ie^{i\varphi/2}\sin\theta/2\end{pmatrix},\quad
\psi_2(\theta,\varphi)=\begin{pmatrix}-ie^{-i\varphi/2} \sin\theta/2\\e^{-i\varphi/2}\cos\theta/2\end{pmatrix},
\]
can be obtained as the linear combination of the eigensolutions of ${\bf D}{\bm\psi}^{(1,2)}_1={\bm\psi}^{(1,2)}_1$. For $N_\theta= N_\varphi=10$ we have that $\psi_1(\theta,\varphi)$ can be obtained from $a_1{\bm\psi}^{(1)}_1+a_2{\bm\psi}^{(2)}_1$, where 
$a_1=9.3217 - 3.1676i$ and $a_2=0.8051 - 0.0499i$. If $\psi_1$ denotes the vector whose elements are the values 
$\psi_1(\theta_j,\varphi_k)$ ordered according to (\ref{ordenlatpun}), the error is given by
\[  
\Vert\psi_1-a_1{\bm\psi}^{(1)}_1-a_2{\bm\psi}^{(2)}_1\Vert_2=8.2\times 10^{-14}.   
\]
Concerning the index, the same simple argument used to show that the continuum Dirac operator has index zero works here: since the lattice Dirac operator is invertible, it has index zero.\\
We end this section by noticing that there exist bosonic-like eigensolutions of the Dirac operator on the sphere. The components of this kind of solutions are $2\pi$-periodic functions in both directions and they can be found through this technique by taking an odd number of nodes $N_\theta$ and $N_\varphi$. The corresponding spectrum consists of zero, with multiplicity 2, and half-integer numbers $\pm 3/2,\pm 5/2,\ldots,\pm (2n+1)/2\ldots$ with multiplicity $2n$.
\section{The Dirac-K\"ahler operator on the sphere}
Let us consider the same parametrization of the unitary sphere given before and let us look for the solution of the Dirac-K\"ahler equation 
\begin{equation}\label{dkem}
D_{\text{DK}}\phi=m\phi,
\end{equation}
where $D_{\text{DK}}=d-\delta$ and $\phi$ is an inhomogeneous differential form on the open chart $(0,\pi)\times(0,2\pi)$. Here, $d$ is the exterior derivative and $\delta$ the adjoint or interior derivative $\delta=*d*$, where $*$ is the Hodge star operator. In the orthonormal basis $(1,d\theta,\sin\theta d\varphi,\sin\theta d\theta\wedge d\varphi)$ of inhomogeneous differential forms, the Dirac-K\"ahler operator takes the form
\begin {equation}\label{dkou}
D_{\text{DK}}=\begin{pmatrix} \text{\large 0}&\begin{matrix}-\displaystyle\frac{\partial}{\partial\theta}-\cot\theta & -\csc^2\theta \displaystyle\frac{\partial}{\partial\varphi}\\
-\displaystyle\frac{\partial}{\partial\varphi} &\displaystyle \frac{\partial}{\partial\theta} \end{matrix}\\ 
\begin{matrix} \displaystyle\frac{\partial}{\partial\theta}  & \csc^2\theta \displaystyle\frac{\partial}{\partial\varphi}\\
\displaystyle\frac{\partial}{\partial\varphi} & -\displaystyle\frac{\partial}{\partial\theta}+\cot\theta  \end{matrix}& \text{\large 0}\\
\end{pmatrix}.
\end{equation}
To obtain a discrete Dirac-K\"ahler operator we proceed as before, with the same definitions of $\Theta$, $D_\theta$ and $D_\varphi$ given in the last section. To find fermionic-like solutions of (\ref{dkem}), i.e., $2\pi$-antiperiodic eigenfunctions in $\theta$ and $\varphi$, we use an even number of nodes $N_\theta$ and $N_\varphi$ in the construction of the differentiation matrices. The numerical solution of the eigenvalue problem of the discrete Dirac-K\"ahler equation produces a two-fold set of real eigenvalues clustered around certain limit values, say $\bar{m}_k$, $k=\pm 1,\pm 2,\cdots,$ to which they converge as $N_\theta$ and $N_\varphi$ grows.  The number of eigenvalues clustered around $\bar{m}_k$ are twice the number corresponding to the multiplicity of eigenvalues of the Dirac equation, and the limit values $\bar{m}_k$ approach the eigenvalues of the Dirac operator on the sphere as $k\to\infty$. The first six positive limit values $\bar{m}_k$ and their distance to the Dirac eigenvalues $k$, are shown in Table 1.\\
\begin{table}[H]\label{tabla}
\begin{center}
\begin{minipage}{12cm}
\caption
{\small The first six positive stabilized eigenvalues of the Dirac-K\"ahler operator on the sphere and their distance to the corresponding eigenvalues of the Dirac operator on $S^2$.}
\end{minipage}\end{center}\vskip.2cm
\hbox to \textwidth{\hfill
\begin{tabular}{|c|c|c|} \hline
\raise 15pt\hbox{\null}\hskip 5pt $k$ \hskip 5pt & $\hskip 1cm \bar{m}_k\hskip 1cm $ & $\hskip 1cm  k-\bar{m}_k \hskip 1cm $\hskip 1cm \\
\hline{\raise 15pt\hbox{\kern-.4em}}
1 & 0.8660254 & 0.1339746\\
2 & 1.9364916 & 0.0635084\\
3 & 2.9580398 & 0.0419602\\
4 & 3.9686269 & 0.0313731\\
5 & 4.9749371 & 0.0250629\\
6 & 5.9791303 & 0.0208697\\
\hline\end{tabular}\hfill}
\end{table}
It is worth to notice that using the basis $\sqrt{\sin\theta}(1\pm i\sin\theta d\theta\wedge d\varphi), \sqrt{\csc\theta}(d\theta \pm i\sin\theta d\varphi)$, the Dirac-K\"ahler operator takes the anti-diagonal form
\begin {equation}\label{dkou}
D_{\text{DK}}=\begin{pmatrix} 0 & i \csc\theta\, D\\ -i \sin\theta\, D & 0\\ \end{pmatrix},
\end{equation}
where $D$ is the Dirac operator on the sphere given in (\ref{diropshp}).\\
Besides the fermionic-like solutions found above, the Dirac-K\"ahler equation have also bosonic-like solutions that can be found by using an odd number of nodes in the construction of $D_\theta$ and $D_\varphi$ to impose $2\pi$-periodic conditions on them. In this case, the eigenvalues are found to be $\pm\sqrt{k(k+1)}$, $k=0,1,\cdots,$ with multiplicity $2(2k+1)$ for $k\ne 0$ and 6 for $k=0$. Since the method yields stable eigenvalues up to 16 digits of precision, i.e. they do not change as $N_\theta$ and $N_\varphi$ grows, the eigensolution in this case should be given in terms of trigonometric polynomials in $\theta$ and $\varphi$.
\\
\section{Final remarks}
The differentiation matrices introduced here to obtain the discrete form of an elliptic operator on a compact manifold have proved to be useful in exploring its spectral problem. This is not an unexpected feature. Examples of compact manifolds are spheres and tori, and on these, some parametrizations involve periodic or antiperiodic conditions on the eigenfunctions which is assumed to be smooth enough in order to be expanded in a Fourier series or approximated by a trigonometric polynomial. Since the $N\times N$ differentiation matrix (\ref{difmattrigpol}) yields the exact derivative of a trigonometric polynomial of degree $(N-1)/2$, a great number of nodes will produce a sufficiently approximate result which becomes the exact solution whenever this solution can be written in terms of trigonometric polynomials. 
\section*{Acknowledgment}
J.L. L\'opez-L\'opez thank CONACYT for financial support  under grant number U50483-F.


\end{document}